\documentclass[prl,twocolumn,
nofootinbib,showpacs,showkeys,floatfix,preprintnumbers,amsmath]{revtex4}
\usepackage{multirow}
\usepackage{esvect}
\usepackage{graphicx}
\usepackage{bm}
\usepackage[T1]{fontenc}
\usepackage{fourier}
\usepackage[utf8]{inputenc}

\begin{document}
%
%
\title{The Fourth Generation Quark and the 750 GeV Diphoton Excess}

\author{Yu-Jie Zhang~$^{1,2}$}
\email{nophy0@gmail.com}
\author{Bin-Bin Zhou~$^{1}$}
\author{Jia-Jia Sun~$^{1}$}
\affiliation{ $^1$~Key Laboratory of Micro-nano Measurement-Manipulation and Physics (
Ministry of Education) and School of Physics,  Beihang University, Beijing 100191, China\\
$^2$~CAS Center for Excellence in Particle Physics, Beijing 100049, China}



\date{ \today}
\begin{abstract} 
Recently, the CMS and ATLAS collaborations have reported a diphoton peak at 750 GeV in the RunII of LHC at 13 TeV.  We assume that  the heavy fourth generation quark doublet $z,~y$ with  $380~$GeV mass, and the width of $z,t$ is much less $b$ quark. 
Then we show that
the contributions of the  $(z \bar z+y \bar y)/\sqrt{2}$ bound state $\eta_{zy}(1S)$ to the diphoton measurements through $\sigma (pp\to \eta_{zy}(1s)\to \gamma \gamma) $ are $  5.6^{+5.6}_{-2.8}~ {\rm fb}$ at  $\sqrt S=13$~TeV. They are constant with the  750 GeV diphoton excess measured by the CMS ant ATLAS collaborations.

\end{abstract}


\maketitle

\section{Introduction}
Recently, the CMS and  ATLAS collaborations have reported a diphoton excess at 750 GeV with width 45 GeV in the RunII of LHC at 13 TeV:
\begin{eqnarray}\label{eq:sigma813}
\sigma(pp\to\gamma\gamma) \approx
\left\{\begin{array}{ll}
(6\pm3){\rm fb}~&\hbox{CMS~\cite{CMS:2015dxe}}\\
(10\pm3){\rm fb}~&\hbox{ATLAS~\cite{date:ATLAS}}.
\end{array}\right.
\end{eqnarray}

A lot of studies have been done here
~\cite{Franceschini:2015kwy,Cao:2015pto,
Bian:2015kjt,Curtin:2015jcv,Fichet:2015vvy,Chao:2015ttq,Demidov:2015zqn,No:2015bsn,
Becirevic:2015fmu,Agrawal:2015dbf,Ahmed:2015uqt,Cox:2015ckc,Kobakhidze:2015ldh,Dutta:2015wqh,Petersson:2015mkr,Low:2015qep,McDermott:2015sck,
Molinaro:2015cwg,Gupta:2015zzs,Ellis:2015oso,Huang:2015svl,
Bi:2015lcf,Chao:2015nac,Cai:2015hzc,Tang:2015eko,Cao:2015apa,Li:2015jwd,Gao:2015igz,
Cao:2015scs,Wang:2015omi,An:2015cgp,Han:2015yjk,Liu:2015yec,Zhang:2015uuo,
Bi:2015uqd,Huang:2015evq,Wang:2015kuj,Cao:2015twy,Ding:2015rxx,Chao:2015nsm,Feng:2015wil,Han:2015dlp,Luo:2015yio,Liao:2015tow}
.
We study the possibility that it is a bound states of the fourth generation quark and anti-quark here. The fourth generation quarks and leptons were proposed about 30 years ago~\cite{Barger:1984jc}. In another hand, the branching fraction $Br(H\to \tau\mu) =(0.84^{+0.39}_{-0.37})\%$ measured by CMS collaboration \cite{Khachatryan:2015kon} may be indicate the fourth generation neutrino here.

In this letter, we assume that, the heavy fourth generate quark $z,~y$ with  $380~$GeV mass and charge $e_z=2/3,~e_y=-1/3$, and the modified CKM matrix elements $|V_{zq}|\sim|V_{yq'}|\ll1$ for $q=d,~s,~b$ and  $q'=u,~c,~t$. Then the width of $z,~y$ quark is much less $b$ quark, it will form the bound state such as  $z\bar u$ or  $z\bar d$, which is long life particle and do not decay in the detector. Then the  bound state  $(z \bar z+y \bar y)/\sqrt{2}$  quarkonium states $\eta_{zy}(1S)$ will be first particle with fourth generation quark component which may be measured at LHC.
The fourth generation quark had been searched by the CMS collaboration at $\sqrt S=8$~TeV, and the lower limit lies between 687 and 782 GeV for all possible values of the branching fractions into the three different final states assuming strong production~\cite{Chatrchyan:2013uxa}. So we require the width of $z,~y$ quark is much less $b$ quark and it will not decay in the detectors.  The bound states of fourth generate quark have been studied in Ref.~\cite{Hagiwara:1990sq,Ishiwata:2011ny,Kim:2015zqa}
In the next section, the decay of $\eta_{zy}(1s)$  is presented, and the production is followed. Finally, a summary is given.

\section{Decay}
First, we study the decay mode of $\eta_{zy}(1s)$ states.
The diphoton and digluon decay widths have been studied to higher order~\cite{Bodwin:1994jh,Guo:2011tz,Li:2012rn,Feng:2015uha}, and the leading order (LO) results are given here,
\begin{eqnarray}
\Gamma(\eta_{zy} \to gg)&=& \frac{16 \alpha_s(m_z)^2 |R_s(0)|^2}{3 m_{\eta_{zy}}^2}, \nonumber \\
\Gamma(\eta_{zy} \to \gamma\gamma)&=&  \frac{6 \alpha(m_z)^2 (e_z^2+e_y^2)^2 |R_s(0)|^2}{m_{\eta_{zy}}^2} \nonumber \\
&=& \frac{50 \alpha(m_z)^2  |R_s(0)|^2}{27 m_{\eta_{zy}}^2},\nonumber \\
\frac{\Gamma(\eta_{zy} \to gg)}{\Gamma(\eta_{zy} \to \gamma\gamma)}&=& \frac{72 \alpha_s(m_z)^2}{25 \alpha(m_z)^2}\sim 4 \times 10^{2},
\end{eqnarray}
where the $|R_s(0)|$ is the radial wave functions of $\eta_{zy}$ at origin.
 And the width of electroweak decay mode of pure $\eta_{zy}$  can be given as,
\begin{eqnarray}
\Gamma(\eta_{zy} \to HH)&=&\Gamma(\eta_{zy} \to H\gamma)  =0,\nonumber \\
\frac{\Gamma(\eta_{zy} \to  f \bar f)}{\Gamma(\eta_{zy} \to \gamma\gamma)}&=&0,
\nonumber \\
\frac{\Gamma(\eta_{zy} \to W^+W^-)}{\Gamma(\eta_{zy} \to \gamma\gamma)}&\sim&2,\nonumber \\
\frac{\Gamma(\eta_{zy} \to ZZ)}{\Gamma(\eta_{zy} \to \gamma\gamma)}&\sim& 3
,\nonumber \\
\frac{\Gamma(\eta_{zy} \to Z\gamma)}{\Gamma(\eta_{zy} \to \gamma\gamma)}&\sim&= 0.5.
\end{eqnarray}
We ignore the ${\cal O}(m_H^2/m_z^2)$, ${\cal O}(m_Z^2/m_z^2)$, ${\cal O}(m_W^2/m_z^2)$, and ${\cal O}(m_{f}^2/m_z^2)$ here.  The $HZ$ and $t\bar t$ widths of $\eta_{z}$ are enhance by $m_z^4/m_Z^4$ and $m_z^2m_t^2/m_Z^4$ respectively,
\begin{eqnarray}
\frac{\Gamma(\eta_{z} \to  HZ)}{\Gamma(\eta_{z} \to \gamma\gamma)}&=&\frac{81 m_z^4}{256 C_w^2S_w^2 m_Z^4}\sim 3\times10^3, \nonumber \\
\frac{\Gamma(\eta_{z} \to  t \bar t)}{\Gamma(\eta_{z} \to \gamma\gamma)}&=&\frac{243 m_z^2m_t^2}{512 C_w^2S_w^2 m_Z^4}\sim 1\times10^3.
\end{eqnarray}
The $HZ$ decay model of $\eta_{z}$ is dominant, which is consistent with Ref.~\cite{Hagiwara:1990sq}.
But if we consider the electroweak decay mode of $\eta_{zy}(\frac{z\bar z+y \bar y}{\sqrt 2})$, we can get
$\Gamma(\eta_{zy} \to  f \bar f)=\Gamma(\eta_{zy} \to HZ)=0$,
which is canceled between $z\bar z$ and $y \bar y$. The ratios of $\frac{\Gamma(\eta_{z} \to VV)}{\Gamma(\eta_{z} \to \gamma\gamma)}$ are given here, 
\begin{eqnarray}
\frac{\Gamma(\eta_{z} \to W^+W^-)}{\Gamma(\eta_{z} \to \gamma\gamma)}&\sim&\frac{81}{128S_w^4\left(1+m_y^2/m_z^2\right)^2}\sim 3,\nonumber \\
\frac{\Gamma(\eta_{z} \to ZZ)}{\Gamma(\eta_{z} \to \gamma\gamma)}&\sim&\frac{\left(32S_w^2-24S_w^2+9\right)^2}{1024C_w^4S_w^4}\sim 0.8
,\nonumber \\
\frac{\Gamma(\eta_{z} \to Z\gamma)}{\Gamma(\eta_{z} \to \gamma\gamma)}&\sim&\frac{\left(8S_w^2-3\right)^2}{32C_w^2S_w^2}\sim 0.2.
\end{eqnarray}

 The electroweak decay width is much less then the strong decay width, and $\Gamma_{tot}(\eta_{zy}(1s))\sim \Gamma(\eta_{zy}(1s)\to gg)$.

The radial wave functions  at origin  $|R_s(0)|$ can be given through the  Schrodinger equation with static  potential. The  gluon exchange potential had been calculated to two loop result~\cite{Peter:1997me,Schroder:1998vy,Kniehl:2004rk}, three loop~\cite{Peter:1996ig,Beneke:2013jia}. We use  one loop potential for $S=0$ states \cite{Gupta:1981pd,Gupta:1982qc} here, where the ${\cal O}(\frac{\alpha_s^2}{m_z})$ potential is included \cite{Melnikov:1998pr,Kniehl:2001ju,Manohar:2000hj,Brambilla:2000gk,Hoang:1998xf,Beneke:1999qg}
\begin{eqnarray}
V_{gluon}(r)&=&-\frac{4\alpha_s}{3r}-\frac{\alpha_s^2}{3\pi r}\left(\frac{31}{3}-\frac{10}{9}n_f\right)-\frac{2\alpha_s^2}{m_zr^2}\nonumber\\
&&
-\frac{4\pi \alpha_s^2}{3m_z^2}\delta{(\vec{r})}+\frac{2 \alpha_s}{3m_z^2r}\left[ \vec{\nabla}^2+\frac{1}{r^2}\vec{r}(\vec{r}\vec{\nabla})\vec{\nabla}\right],\nonumber
\end{eqnarray}
where  $n_l=5$ is the number of light-quark flavors. 
The photon exchange potential is 
\begin{eqnarray}
V_{photon}(r)&=&- \frac{5}{18} \frac{ \alpha}{ r},
\end{eqnarray}
The Higgs exchange potential had been calculated to one loop, where the large fermion mass is canceled out and hence produces no enhancement for the radiative corrections~\cite{Kuchiev:2010ia,Kuchiev:2010hz}. So that the Higgs exchange  potential in the momentum space is given as \cite{Strassler:1990nw,Ishiwata:2011ny}:
\begin{eqnarray}
V_{Higgs}(r)&=&-  \frac{g_{z}^2e^{-m_Hr}}{4\pi r}\left(1+\frac{m_z^2}{2\pi^2v^2}+\frac{7m_t^2}{16\pi^2v^2}\right),
\end{eqnarray}
where 
$g_z=\frac{ m_z}{v}$, $\frac{g_{z}^2}{4\pi} \sim 0.2$, $v=246~{\rm GeV}$, and $m_H=125~{\rm GeV}$.

The  relativistic corrections of kinetic energy and Higgs exchange potential and the contributions of ghost of $W,~Z$ are~\cite{Ishiwata:2011ny}
\begin{eqnarray}
&&V_{rel.}(r)=-\frac{(\vec{\nabla}^2)^2}{4m_z^3}-\frac{1}{2v^2}\delta{(\vec{r})}\nonumber \\
&&~~~~~~~~~+\frac{m_H^2e^{-m_Hr}
-m_Z^2e^{-m_Zr}
+2m_W^2e^{-m_Wr}}{16\pi v^2 r}.
\end{eqnarray}
The first term are the relativistic corrections of kinetic energy.
The total potential is
\begin{eqnarray}
V(r)=V_{gluon}(r)+V_{photon}(r)+V_{Higgs}(r)+V_{rel.}(r)
\end{eqnarray}

In the numerical calculation, the strong coupling constant can be determined through  $\alpha_s(m_z v_z/2) \sim v_z$ in the non-relativistic bound states\cite{Bodwin:1994jh}, then $\alpha_s\sim 0.162$. The variational method is used here, and the test function is select as the hydrogen atom radial wave functions
\begin{eqnarray}
\label{eq:testWF}
R_{1s}(r)=2\times q^{3/2}\times e^{-q r}.
\end{eqnarray}

Then we get 
the radial wave functions of $\eta_{zy}(1S)$ is
\begin{eqnarray}
q&=&98~{\rm GeV}\nonumber \\
\frac{|R_{1s}(0)|^2}{m_{\eta_{zy}}^2}&=&6.9 ~{\rm GeV}^3
.
\end{eqnarray}
If we consider the LO Coulomb potential only, $\frac{|R_{1s}(0)|^2}{m_{\eta_{zy}}^2}\sim0.1~{\rm GeV}$, wihch is consistent with the result in~\cite{Hagiwara:1990sq}. If we consider the LO Coulomb   and higgs exchange potential  only, our result is consistent with  the result in~\cite{Hagiwara:1990sq,Kim:2015zqa}. The other potential terms absorb the wavefunction to the origin and enhance $q$  in Eq.(\ref{eq:testWF}).

In another hand, $\alpha_s (m_z)\sim 0.092$, $\alpha (m_z)\sim 1/120$.
We can get the properties of $\eta_{zy}(1S)$  states~\cite{Bodwin:1994jh},
\begin{eqnarray}
E_1&=&-10~{\rm GeV}, \nonumber \\
M_{\eta_{zy}(1s)}&=&2m_z+E_1=750~{\rm GeV} , \nonumber \\
\Gamma(\eta_{zy}(1s) \to gg)&=& 0.30~{\rm GeV}, \nonumber \\
\Gamma(\eta_{zy}(1s) \to \gamma\gamma)&=&  0.86~{\rm MeV},\nonumber \\
Br(\eta_{zy}(1s) \to \gamma\gamma)&=& 2.8 \times 10^{-3}.
\end{eqnarray}

\section{Production}
The leading order (LO) parton cross section of $\eta_{zy}(1s)$ can be get through \cite{Spira:1995rr}:
\begin{eqnarray}
\hat\sigma_{LO} (gg\to \eta_{zy}(1s)) & = & {\sigma_0}{M_{\eta_{zy}(1s)}^2} \delta
(\hat s -M_{\eta_{zy}(1s)}^2)
\end{eqnarray}
and
\begin{eqnarray}
\sigma_0 & = & \frac{\pi^2}{8 M_{\eta_{zy}(1s)}^3} \Gamma_{LO} (\eta_{zy}(1s) \to gg)
\end{eqnarray}
where $\hat{s}$ is the invariant energy squared of initial states $gg$.
And the hadronic cross section of $\eta_{zy}(1s)$ is
\begin{eqnarray}
&&\sigma_{LO} (pp\to \eta_{zy}(1s))\nonumber \\
 & = &  \sigma_0 \tau \int_{\tau}^{1} \frac{dx}{x} f(x)f(\frac{\tau}{x}) \nonumber \\
& = &  \frac{\Gamma_{LO} (\eta_{zy}(1s) \to gg)}{ M_{\eta_{zy}(1s)}S}  \frac{\pi^2}{8}\int_{\tau}^{1} \frac{dx}{x} f(x)f(\frac{\tau}{x}),
\end{eqnarray}
where $\tau=M_{\eta_{zy}(1s)}^2/S$, and $S$ is the invariant mass square of initial proton-proton. Then its contribution to the diphoton distribution is \begin{eqnarray}
&&\sigma_{LO} (pp\to \eta_{zy}(1s)\to \gamma\gamma)\nonumber \\
& = &  \frac{\Gamma_{LO} (\eta_{zy}(1s) \to \gamma\gamma)}{ M_{\eta_{zy}(1s)}S}  \frac{\pi^2}{8}\int_{\tau}^{1} \frac{dx}{x} f(x)f(\frac{\tau}{x}),
\end{eqnarray}
Then $ \frac{\pi^2}{8}\int_{\tau}^{1} \frac{dx}{x} f(x)f(\frac{\tau}{x})$ 
is 2137 for   $\sqrt S=13$~TeV~\cite{Liao:2015tow,Franceschini:2015kwy}.
Then we can get the cross sections at $\sqrt S=13$~TeV,
\begin{eqnarray}
\sigma (pp\to \eta_{zy}(1s)\to \gamma \gamma) & = & 5.6^{+5.6}_{-2.8}~ {\rm fb}.
\end{eqnarray}
The relative  error bar is estimated as $1^{+1.0}_{-0.5}$ from the $\alpha_s$, factorization scale, higher order corrections, and so on.
The higher states $\eta_{zy}(ns)$ for $n=2,~3,...$ will contribution to the diphoton distribution, which may enlarge the measured width of the diphoton excess.

\section{Summary}
In summary,  we assume that  the heavy fourth generation quark doublet $z,~y$ with  $380~$GeV mass, and the width of $z,t$ is much less $b$ quark. 
Then we show that
the contributions of the  $(z \bar z+y \bar y)/\sqrt{2}$ bound state $\eta_{zy}(1S)$ to the diphoton measurements through $\sigma (pp\to \eta_{zy}(1s)\to \gamma \gamma) $ are $  5.6^{+5.6}_{-2.8}~ {\rm fb}$ at  $\sqrt S=13$~TeV. They are constant with the  750 GeV diphoton excess measured by the CMS ant ATLAS collaborations.

%
%
%

\begin{acknowledgments}
We would like to thank K.T. Chao, B.Q. Li for useful discussions.
This work is supported by the National Natural Science
Foundation of China (Grants No. 11375021,11575017), the New Century Excellent Talents in University (NCET) under grant
NCET-13-0030,  the Major State Basic Research Development Program of China (No. 2015CB856701),  and
the Fundamental Research Funds for the Central Universities.

\end{acknowledgments}


\providecommand{\href}[2]{#2}\begingroup\raggedright\endgroup

\end{document}